\documentclass[aps,prl,twocolumn,superscriptaddress,showpacs]{revtex4}

\usepackage{graphicx}
\usepackage{epsfig}
\usepackage[english]{babel}
\usepackage{amsbsy}
\makeatother

\def\lsim{\mathrel{\rlap{\lower4pt\hbox{\hskip1pt$\sim$}}
    \raise1pt\hbox{$<$}}}                
\def\gsim{\mathrel{\rlap{\lower4pt\hbox{\hskip1pt$\sim$}}
    \raise1pt\hbox{$>$}}}                
\begin{document}

\title{Evidence for the Production of Slow  Antiprotonic Hydrogen in Vacuum}

\author{N. Zurlo} 
\affiliation{Dipartimento di Chimica e Fisica per l'Ingegneria e per i Materiali,
 Universit\`a di Brescia, 25123 Brescia, Italy} 
\affiliation{Istituto Nazionale di Fisica Nucleare, Gruppo Collegato di Brescia, 25123 Brescia, Italy} 

\author{M. Amoretti}
\affiliation{Istituto Nazionale di Fisica Nucleare, Sezione di Genova, 16146 Genova Italy} 

\author{C. Amsler}
\affiliation{Physik-Institut, Z\"urich University, CH-8057 Z\"urich, Switzerland}

\author{G. Bonomi} 
\affiliation{Dipartimento di Ingegneria Meccanica,
 Universit\`a di Brescia, 25123 Brescia, Italy} 
\affiliation{Istituto Nazionale di Fisica Nucleare, Universit\`a di Pavia, 27100 Pavia, Italy} 


\author{C. Carraro}

\affiliation{Istituto Nazionale di Fisica Nucleare, Sezione di Genova, 16146 Genova Italy} 
\affiliation{Dipartimento di Fisica, Universit\`a di Genova, 16146 Genova, Italy} 

\author{C. L. Cesar}
\affiliation{Instituto de Fisica, Univesidade Federal do Rio de Janeiro, Rio de Janeiro 21945-970}

\author{M.~Charlton}
\affiliation{Department of Physics, University of Wales Swansea, Swansea SA2 8PP, UK} 

\author{M.~Doser}
\affiliation{EP Division, CERN, CH-1211 Geneva 23, Switzerland} 


\author{A.~Fontana} 
\affiliation{Istituto Nazionale di Fisica Nucleare, Universit\`a di Pavia, 27100 Pavia, Italy} 
\affiliation{Dipartimento di Fisica Nucleare e Teorica, Universit\`a di Pavia, 27100, Pavia, Italy} 


\author{R. Funakoshi}
\affiliation{Department of Physics, University of Tokyo, Tokyo 113-0033, Japan} 

\author{P.~Genova}
\affiliation{Istituto Nazionale di Fisica Nucleare, Universit\`a di Pavia, 27100 Pavia, Italy} 
\affiliation{Dipartimento di Fisica Nucleare e Teorica, Universit\`a di Pavia, 27100, Pavia, Italy} 


\author{R. S. Hayano}
\affiliation{Department of Physics, University of Tokyo, Tokyo 113-0033, Japan} 

\author{L.~V.~J{\o}rgensen}
\affiliation{Department of Physics, University of Wales Swansea, Swansea SA2 8PP, UK} 

\author{A.~Kellerbauer}
\affiliation{EP Division, CERN, CH-1211 Geneva 23, Switzerland} 

\author{V.~Lagomarsino}
\affiliation{Istituto Nazionale di Fisica Nucleare, Sezione di Genova, 16146 Genova Italy}
\affiliation{Dipartimento di Fisica, Universit\`a di Genova, 16146 Genova, Italy} 

\author{R. Landua} 
\affiliation{EP Division, CERN, CH-1211 Geneva 23, Switzerland} 

\author{E. Lodi Rizzini}
\affiliation{Dipartimento di Chimica e Fisica per l'Ingegneria e per i Materiali, Universit\`a di Brescia, 25123 Brescia, Italy} 
\affiliation{Istituto Nazionale di Fisica Nucleare, Gruppo Collegato di Brescia, 25123 Brescia, Italy} 

\author{M. Macr\`\i}
\affiliation{Istituto Nazionale di Fisica Nucleare, Sezione di Genova, 16146 Genova Italy} 

\author{N.~Madsen}
\affiliation{Department of Physics, University of Wales Swansea, Swansea SA2 8PP, UK} 

\author{G. Manuzio}
\affiliation{Istituto Nazionale di Fisica Nucleare, Sezione di Genova, 16146 Genova Italy}
\affiliation{Dipartimento di Fisica, Universit\`a di Genova, 16146 Genova, Italy} 

\author{D.~Mitchard}
\affiliation{Department of Physics, University of Wales Swansea, Swansea SA2 8PP, UK} 

\author{P. Montagna}
\affiliation{Istituto Nazionale di Fisica Nucleare, Universit\`a di Pavia, 27100 Pavia, Italy}
\affiliation{Dipartimento di Fisica Nucleare e Teorica, Universit\`a di Pavia, 27100, Pavia, Italy}

\author{L.G. Posada}
\affiliation{Department of Physics, University of Tokyo, Tokyo 113-0033, Japan}

\author{H. Pruys}
\affiliation{Physik-Institut, Z\"urich University, CH-8057 Z\"urich, Switzerland}

\author{C. Regenfus} 
\affiliation{Physik-Institut, Z\"urich University, CH-8057 Z\"urich, Switzerland}

\author{A.~Rotondi}
\affiliation{Istituto Nazionale di Fisica Nucleare, Universit\`a di Pavia, 27100 Pavia, Italy}
\affiliation{Dipartimento di Fisica Nucleare e Teorica, Universit\`a di Pavia, 27100, Pavia, Italy}

\author{G. Testera}
\affiliation{Istituto Nazionale di Fisica Nucleare, Sezione di Genova, 16146 Genova Italy} 

\author{D.P. Van der Werf} 
\affiliation{Department of Physics, University of Wales Swansea, Swansea SA2 8PP, UK} 

\author{A. Variola}
\affiliation{Istituto Nazionale di Fisica Nucleare, Sezione di Genova, 16146 Genova Italy} 

\author{L. Venturelli}
\affiliation{Dipartimento di Chimica e Fisica per l'Ingegneria e per i Materiali, Universit\`a di Brescia, 25123 Brescia, Italy}
\affiliation{Istituto Nazionale di Fisica Nucleare, Gruppo Collegato di Brescia, 25123 Brescia, Italy} 

\author{Y. Yamazaki}
 \affiliation{Atomic Physics Laboratory, RIKEN, Saitama 351-0198, Japan} 

\affiliation{}

\collaboration{ATHENA Collaboration}
\noaffiliation

\date{\today}

\begin{abstract}
We present evidence showing how antiprotonic hydrogen, the quasi-stable antiproton ($\bar p$)-proton bound system, has been synthesized following the interaction of antiprotons with the molecular ion H$_2^+$ in a nested Penning trap environment. From a careful analysis of the spatial distributions of antiproton annihilation events, evidence is presented for antiprotonic hydrogen production with sub-eV kinetic energies in states around $n$ = 70, and with low angular momenta. The slow antiprotonic hydrogen may be studied using laser spectroscopic techniques.
\end{abstract}


\maketitle

Studies of the properties of the two-body hydrogenic bound states of the stable leptons and baryons have produced some of the most precise measurements of physical quantities and provided powerful tests of our understanding of the laws of nature. Interest in this area is still strong, following the recent production of antihydrogen,  $\mathrm{\bar H}$, at low energies \cite{Amoretti:2002um,Gabrielse:2002}. Accurate comparisons of the transitions in hydrogen and antihydrogen are eagerly awaited as a stringent test of CPT symmetry.

Antiprotonic hydrogen ($\bar p p$) is also of interest. 
 Its level structure is similar to that of hydrogen, but with much larger
 binding energies. Precision measurements of its spectroscopic properties may
 allow determination of the so-called antiprotonic Rydberg constant and/or the
 antiproton/electron mass ratio.

Although $\bar p p$  has been studied extensively in the past, this has exclusively been achieved
 by stopping antiprotons in liquid or gaseous targets for X-ray spectroscopy of inner shell cascades,
 or for the production of new light mesons and baryons (see e.g. \cite{Batty:1989,Montanet:2001} ).
Here we report a radically new method of $\bar p p$ production resulting in emission almost at rest in vacuum. This has been achieved using a chemical reaction between antiprotons and molecular hydrogen ions (H$_2^+$) in the ATHENA Penning trap apparatus \cite{Amoretti:2002um, Amoretti:2004uf}. This advance has opened the way for laser spectroscopic studies of  $\bar p p$, or other antiprotonic systems formed by $\bar p$ interactions with HD$^+$ or D$_2^+$, akin to those successfully deployed in the study of antiprotonic helium (see e.g. \cite{Yamazaki:2002,Hori:2005}).

The experiments were made possible by the availability of a high-quality low
energy $\bar p$ beam delivered by the CERN Antiproton Decelerator to the ATHENA
$\bar\mathrm{H}$ apparatus. The latter contained a multi-electrode system of cylindrical Penning traps, 2.5 cm in diameter and $\sim 90$~cm in length immersed in an axial magnetic field of 3 T. The residual pressure of $\sim 10^{-12}$ Torr, in the 15 K cryogenic environment of the trap, was due to hydrogen and helium gases. The central region contained the mixing trap: a nested Penning trap, approximately  10 cm long, that allowed positrons, $e^+$, and antiprotons  to be confined simultaneously. For $\bar \mathrm{H}$ production the mixing trap contained a spheroidal cloud of $\sim 3.5 \times 10^7~e^+$. Around $10^4$~antiprotons were injected into this plasma, with the resulting $\bar p$ annihilations monitored for 60 s by position sensitive detectors \cite{Amoretti:2004uf,Regenfus:detector}.
 These registered the passage of the charged pions, to localize annihilation
 vertices which were due, not only to $\bar \mathrm{H}$ formation followed by
 annihilation on the electrode surface \cite{Amoretti:2002um, Amoretti:2004ci},
 but also $\bar p$ annihilation following transport to the electrode walls
 \cite{Fujiwara:2004ra}, and annihilation following interactions with residual
 gas atoms or ions present in the trap. It was shown in \cite{Amoretti:2002um,
 Amoretti:2004ci} that the vertex data were predominantly $\bar \mathrm{H}$
 annihilations during so-called ``cold mixing'' (CM), when the $e^+$ cloud was
 held at the trap ambient of 15 K. In contrast for ``hot mixing'' (HM), when the $e^+$
 were heated to a temperature, $T_e$, of several thousand K (here 8000 K)
 \cite{amoretti:modi, amoretti:modii}, $\bar \mathrm{H}$ formation was
 suppressed and the $\bar p$ annihilations were mainly a result of collisions
 with trapped positive ions. It is this effect (which is also present for CM) that is addressed here.

\begin{figure}[t!]
\epsfig{file=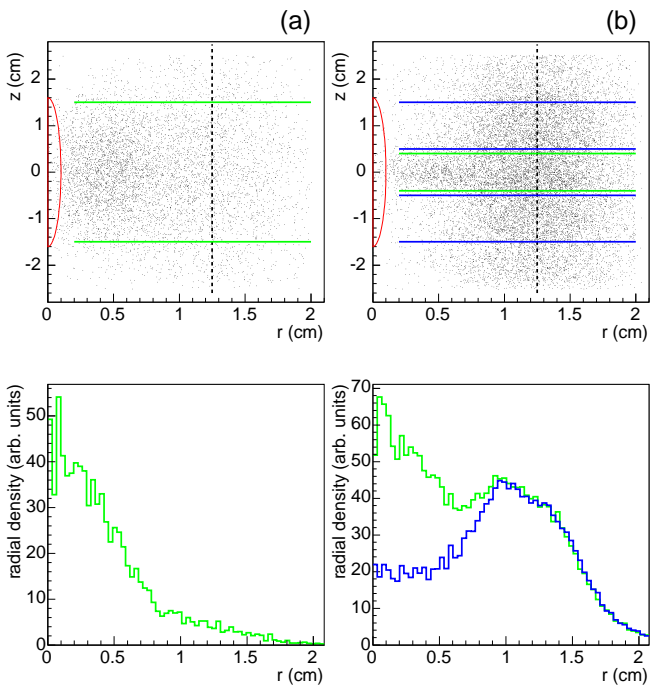,width=\columnwidth}
\caption{$r-z$ 
scatter plot and radial densities $\left(\frac{1}{r}\frac{d N}{d r}\right)$ of the annihilation vertices for: 
(a) hot mixing; (b) cold mixing. The dashed black line indicates the position of the trap wall; the red semi-ellipse shows the section of the $e^+$ plasma. The green radial densities are for the corresponding central $z$-region events (inside the green lines) whilst the data in blue are for the 2 lateral $z$-regions, normalized for $r>1.25$~cm.}
\label{figura1}
\end{figure}

Fig.\ref{figura1} shows $r-z$ scatter plots for annihilation vertices taken
under HM and CM conditions. Here the radial positions, $r$, (i.e. the distance from the trap axis) of the events are plotted versus their axial coordinates, $z$. Also shown are the attendant radial density distributions $\left(\frac{1}{r}\frac{d N}{d r}\right)$. The distributions are broadened by the uncertainties in the vertex determination (around 1.8~mm in the $z$-direction and 3.5~mm in the transverse dimensions) caused mainly by the inability of the ATHENA detector to reconstruct the curved trajectories of the pions in the 3 T magnetic field. The present HM results are for the highest $T_e$ achieved by ATHENA.

There are striking differences between the two $r-z$ plots. Besides the $\bar
\mathrm{H}$ annihilations on the trap wall centred around $r = 1.25$~cm (CM
only), there are events localized at smaller radii which dominate in HM
(Fig.\ref{figura1}a), but which are also evident in CM (Fig.\ref{figura1}b).
Examining Fig.\ref{figura1}b, the radial density distributions for CM at small radii ($r\lesssim 0.5$~cm) behave quite differently between the central ($|z|<0.5$~cm) and its adjacent regions ($\rm 0.5~cm < |z| < 1.5~cm$). Moreover, the shape of the distribution for the events in the central region resembles that for HM.

\begin{figure}
\epsfig{file=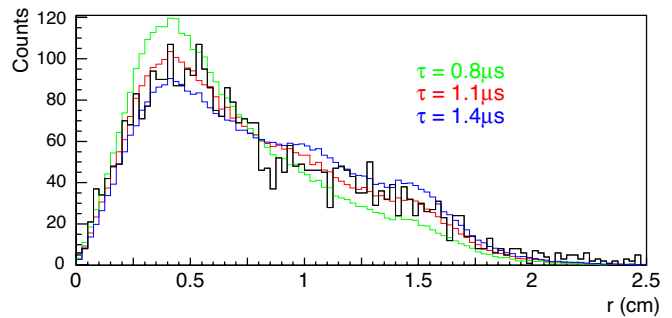,width=\columnwidth}
\caption{Experimental radial distribution of $\bar p$ annihilation vertices {(black histogram)} for HM ($-$1.5~cm$<z<1.5$~cm) with a Monte Carlo simulation ($T_e = 8000$~K, $v_{th}$ = 5600~ms$^{-1}$, generation on the surface of a spheroid with $z_p=16$~mm and $r_p = 1$~mm rotating with a frequency of 300~kHz, i.e. $v_{tang}=2000$~ms$^{-1}$); see text for details.
Results of simulations with different mean lifetimes are shown:
green, $\tau= 0.8 \mu$s ($\chi^2_{red}=2.78$);
 red,  $\tau= 1.1 \mu$s ($\chi^2_{red}=1.48$);
 blue, $\tau= 1.4 \mu$s ($\chi^2_{red}=2.14$).}
\label{figura2}
\end{figure}

\begin{figure}
\epsfig{file=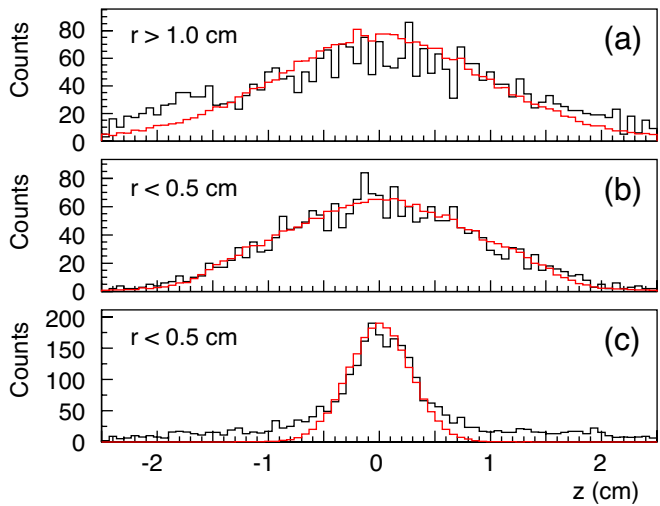,width=\columnwidth}
\caption{Experimental axial distributions: hot mixing for events near the trap wall (a), and near the trap axis (b) and for cold mixing for events near the trap axis (c). In (a,b): the red line is the simulation with the parameters of 
Fig.\ref{figura2} and with lifetime $\tau=1.1\mu$s.
In (c) the red line is the simulation with the parameters of
Fig.\ref{figura4}a.}
\label{figura3}
\end{figure}

In Fig.\ref{figura2} and Figs.\ref{figura3}a,b the radial $\left(\frac{d N}{d r}\right)$ and axial $\left(\frac{d N}{d z}\right)$ annihilation
distributions are plotted for the HM case. Fig.\ref{figura3}c shows the
axial distribution for CM for events with $r<0.5$~cm being notably narrower than the HM case in Fig.\ref{figura3}b. 

In order to determine the characteristics of the near-axis events the possibility that they are due to $\bar \mathrm{H}$ has been investigated. In ATHENA, $\bar \mathrm{H}$ was detected by the coincidence in space and in time (within 5 $\mu$s) of $\bar p$ and $e^+$ annihilations. This was achieved for events having a charged pion vertex
accompained by a pair of 511 keV photons with the angle between them denoted $\theta_{\gamma\gamma}$
\cite{Amoretti:2002um,Amoretti:2004ci}.
Examination of the distributions of $\cos\theta_{\gamma\gamma}$ indicates that
the trap centre events are not due to $\bar \mathrm{H}$, except for a small
fraction in CM where the long tails in Fig.\ref{figura3}c are due to poor
reconstructions of $\bar \mathrm{H}$ annihilations on the trap wall. No
$\bar\mathrm{H}$ annihilations occur for HM. Moreover,
$\cos\theta_{\gamma\gamma}$ distributions for CM in the three $z$-regions in Fig.\ref{figura1}b suggest that the fraction of non-$\bar \mathrm{H}$ annihilations on the wall in the central $z$-region is insignificant.

The distributions in Figs.\ref{figura1}--\ref{figura3} also show that in HM the annihilation distribution is extended and has a conspicuous fraction out to the trap wall. We interpret these features as evidence for the production of slow antiprotonic hydrogen as follows:

\begin{enumerate}
\item[i)] the limited axial range of the vertices, when compared to the total length of the nested trap, indicates that $\bar p$ do not annihilate in-flight on residual gas which is present throughout;
\item[ii)] the events cannot correspond to in-flight annihilation on trapped positive ions since the latter are only present near the $e^+$ cloud whilst the annihilation events are radially diffuse;
\item[iii)] positive ions can capture $\bar p$ in the central part of the recombination trap.
In the case of $\bar p \mathrm{He^+}$ the residual electron would be rapidly ejected, in the majority of cases in less than 10 ns \cite{korobov:2003}, leaving a charged system that would annihilate very near its point of formation. Since this is inconsistent with our observations, $\bar p$ capture by helium ions can be excluded.
\end{enumerate}

Further information on the inferred antiprotonic system formed from capture by a positive ion can be obtained by exploiting the different charged pion multiplicities expected for $\bar  p$ annihilation on a proton or neutron. Tab.\ref{tabella1} shows the ratios, $R_{23}$, of the number of the reconstructed annihilation vertices having two charged pion tracks to those with three tracks for different data samples.

\begin{table}[hbt]
\begin{tabular}{|l|c|c|}
\hline
Data set &
Ratio $R_{23}$ on wall&
Ratio $R_{23}$ at centre\tabularnewline
\hline
\hline
Cold mixing&
1.35$\pm$0.01&
1.22$\pm$0.04\tabularnewline
\hline
Hot mixing&
1.38$\pm$0.10&
1.17$\pm$0.04\tabularnewline
\hline
{antiprotons}  only&
1.40$\pm$0.03&
\tabularnewline
\hline
\hline
Monte Carlo $\bar pp$&
$1.19\pm 0.01$&
$1.19\pm 0.01$
\tabularnewline
\hline
\end{tabular}
\caption{Experimental and Monte Carlo results for the number of
charged pion tracks due to $\bar p$ annihilations. }
\label{tabella1}
\end{table}

From Tab.\ref{tabella1}, the $R_{23}$ values for annihilation on the trap wall for all the samples agree, within uncertainties, but differ from those for the trap centre by 4 standard deviations. This is not due to geometry, since the Monte Carlo simulations assuming the $\bar p p$ system give the same result both for the ``wall'' and ``centre'' annihilations. It is likely that the trap centre events are due to $\bar pp$ annihilation, since the $\bar p p$ Monte Carlo result agrees well with the experiment. This, together with the three constraints described above, suggest that  $\bar p p$ is responsible for the observed annihilation distributions. Thus, the data can be further analysed to search for consistency with the conditions pertaining to the production region at the two different positron temperatures, the  $\bar p p$  lifetime and finally the energetics of the formation reaction.

First,  $\bar p p$  is not confined by the electromagnetic fields of the traps and if formed in a metastable state \cite{Reifenrother:1989nq, Batty:1989, Cohen:2004, Sakimoto:2004} it can decay in flight far from its point of formation, perhaps even on the trap wall. The convolution of the  $\bar p p$  lifetimes and velocities governs the observed annihilation distributions. Moreover, Figs. \ref{figura3}b and c clearly indicate that the temperature of the $e^+$ cloud influences the spatial origin of the  $\bar p p$. 

We have attempted to generate the initial position of the  $\bar p p$ using a Monte Carlo simulation which takes into account the radius, $r_p$, and the axial half-length, $z_p$, of the $e^+$ spheroid, both deduced by means of the non-destructive technique described in \cite{amoretti:modi, amoretti:modii}. Parameters obtained using this technique allow the plasma rotation frequency to be extracted.

In this simulation, the spheroid was characterized by $r_p=1$~mm and $z_p=16$~mm. For the CM case the $\bar p p$ was generated in a region with a fixed radial position at $r=r_p=1$~mm and with a Gaussian distribution along the axis centred at the symmetry plane of the plasma with $\sigma=2.5$~mm. This gave the best fit to the data.  For the HM case, however,  $\bar p p$ was generated with $\sigma=10$~mm, though limited to the length of $e^+$ plasma.
It is notable that, for the HM case, the simulated annihilation
distributions were not strongly dependent upon the assumed starting
conditions, taking into account our experimental resolution.


The velocity of the  $\bar p p$  was generated from the sum of a thermal Maxwellian
distribution, $v_{th}$, and the tangential velocity, $v_{tang}$, induced by the
$\mathrm{\vec{E} \times \vec{B}}$ plasma rotation as $v_{tang}= \mathrm{\vec{E}
\times \vec{B}/|B|^2}$. Following this prescription, the mean radial kinetic
energy of the $\bar pp$ is about 40 meV for CM (15 K), and dominated by the effect of the plasma rotation, and about 700 meV in the HM case (8000 K), and dominated by the plasma temperature. An exponential decay law for the  $\bar p p$  lifetime distribution was assumed such that its mean lifetime was determined by fitting the simulations to the observed data.

The simulated radial and axial annihilation distributions are plotted with the
HM data in Fig.\ref{figura2} and Fig.\ref{figura3}a,b. The agreement is good and the best fit was obtained with a mean lifetime of (1.1 $\pm$ 0.1)$\mu$s. The sensitivity of the fit to this lifetime is illustrated in Fig.\ref{figura2} by the clear discord between the experimental data and the simulations when lifetimes of 0.8 $\mu$s and 1.4 $\mu$s were used in the latter. The fits imply that about 25\% of the  antiprotonic hydrogen  atoms annihilate on the trap wall in HM.

\begin{figure}
\epsfig{file=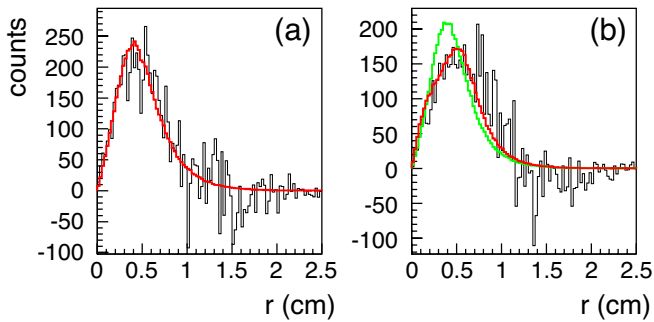,width=\columnwidth}
\caption{Experimental radial distribution
for cold mixing 2003 (a) and cold mixing 2002 (b), obtained by subtracting the
$\bar \mathrm{H}$ contribution (see text). The red lines are
the simulation results. In (a): T=15 K, $v_{th} = 250$~ms$^{-1}$,
generation from a spheroid with $z_p= 16$~mm and $r_p = 1$~mm rotating with  a frequency of 300~kHz, i.e. $v_{tang}=2000$~ms$^{-1}$; mean lifetime is $\tau$= 1.1~$\mu$s. In (b): same parameters as in (a) except $r_p = 2.5$~mm and the rotation frequency  is 80~kHz, i.e. $v_{tang}=1300$~ms$^{-1}$. In (b) the green line corresponding to the parameters of the red line in (a) is shown for comparison.}
\label{figura4}
\end{figure}

To isolate the radial distribution of the non-$\bar \mathrm{H}$ annihilations in
the central $z$-region for the CM sample the normalized distribution, as
evaluated from Fig.\ref{figura1}b for the two lateral $z$-regions, was
subtracted from the central one. Since, as noted above, non-$\bar \mathrm{H}$
annihilations on the wall in the central $z$-region are insignificant for CM, the radial distributions were normalised for $r>1.25$~cm.

The results are plotted in Fig.\ref{figura4}a. The Monte Carlo simulated events,
assuming the same lifetime extracted from fits to the HM data, also show
good agreement for CM. In this case, the simulation indicates that $< 0.5\%$ of the $\bar p p$ reaches the trap surface.

A further test of our model has been obtained by examining a sample of CM data
acquired by ATHENA in 2002 with a $e^+$ plasma radius of 2.5 mm (and hence a
different rotation frequency). Fig.\ref{figura4}b shows the radial vertex
distribution of this sample together with Monte Carlo simulations for two values
of the radius of the $\bar p p$ source. The simulated events for the 2.5 mm source are
in much better agreement with the experimental data than those for 1.0 mm. This
is illustrative of the sensitivity of the model to the spatial origin of the $\bar p p$ 
for CM samples.

A possible explanation of the experimental differences between the  $\bar p p$  distributions in CM and HM 
lies in the nature of the thermal equilibrium state of the combination of a $e^+$ plasma with an admixture of ions. The physics of the radial separation of the different species in two-component plasmas has been given elsewhere \cite{Dubin:1999, Oneil:1981} and has been experimentally observed for a mixture of positrons and ${}^9$Be$^+$ in \cite{jelenkovic:2003}. For our experimental conditions, assuming thermal equilibrium, the centrifugal potential barrier is of the order of 10 meV. Thus, for CM at 15 K, the thermal energy of the ions means that they will be partially separated from the $e^+$ and confined near the equatorial region of the plasma.  However, at a $e^+$ temperature of 8000~K the barrier is negligible and the ions will be present throughout the plasma.

We note here that the experimental data cannot be reproduced by simulation if it is assumed that the  $\bar p p$ gains a recoil energy of the order of 1 eV or higher. Thus, it is contended that  $\bar p p$ is being produced in a recoil-free collision of an antiproton with a positive ion such that its velocity is determined predominantly by the thermal and plasma environment. This insight, together with the constraints i)-iii) noted above, suggest that the only possible collision partner for the $\bar p$ is the molecular ion, $\mathrm{H_2^+}$. Thus, the inferred $\bar p p$ production mechanism is,
\begin{equation}
\label{eq1}
\bar{p}+ \mathrm{H^+_2}\rightarrow  {\bar p p}(n,l) +\mathrm{H}.
\end{equation}

Hence we believe that ATHENA has observed around 100 antiprotonic hydrogen annihilations every 60 s $\bar p$ injection
cycle for the CM and HM conditions. In an experiment in which the number of ions
present in our nested well was counted by measuring the charge collected
following emptying of the trap it has been deduced that the trap contained H$^+_2$ ions. 
These probably arose as a result of the positron loading procedure \cite{jorgensen},
 in which  the positrons could collide
with H$_2$ residual gas as they were slowly squeezed into the mixing region.
Ions may also be produced and trapped during $\bar p$ loading (see also \cite{Gabrielse:1999}) and 
we estimate that around $10^4- 10^5$ ions were present under typical ambient conditions.
 It is straightforward to show that this ion density, together with the observed  $\bar p p$ production rate and the $\bar p$ speeds used in the simulation are consistent with calculated cross sections for reaction (1) \cite{ Sakimoto:2004}. 



Agreement with \cite{ Sakimoto:2004} does not, unfortunately, extend to the most
likely  principal quantum numbers of the antiprotonic hydrogen atoms produced in the reaction. The calculation
\cite{ Sakimoto:2004} finds production peaked around $n$ = 34, in the presence
of substantial  $\bar p p$ recoil. The latter is contrary to observation. Simple
kinematics relating to near zero-energy $\bar p-\mathrm{H}_2^+$ collisions
suggest that $n$ = 68 should dominate, with the liberated hydrogen atom in its
ground state. In this case the lifetime of 1.1 $\mu$s extracted from the
simulations implies that production in low angular momentum states ($l < 10$) is favoured, since the radiative lifetime to an $l$ = 0 state (which is followed by prompt annihilation) weighted by the statistical ($2l + 1$) distribution is around 15 $\mu$s for $n$ = 68. The dominance of low $l$ is intuitive for such a slow collision in which the molecular ion will be severely polarised by the incoming $\bar p$, resulting in an almost collinear collision system.

In conclusion we have presented evidence for the production of antiprotonic hydrogen, in vacuum, with sub-eV kinetic energies and in a metastable state. Given the capability of accumulating $10^8$ H$_2^+$ ions in tens of seconds and storing $\sim 5 \times 10^6~$ antiprotons in some minutes \cite{Yamazaki:2005, Oshima:2004}, our result opens up the possibility of performing detailed spectroscopic measurements on antiprotonic hydrogen as a probe of fundamental constants and symmetries.

This work was supported by CNPq, FAPERJ, CCMN/UFRJ (Brazil), INFN (Italy), MEXT (Japan), SNF (Switzerland), SNF (Denmark) and EPSRC (UK).


\end{document}